\documentclass[natbib,preprint2]{aastex}

\newcommand{\AlIII}{\mbox{Al\,{\sc iii}}}
\newcommand{\CIV}{\mbox{C\,{\sc iv}}}

\newcommand{\feka}{\mbox{Fe\,K$\alpha$}}

\newcommand{\MgII}{\mbox{Mg\,{\sc ii}}}

\newcommand{\kms}{\mbox{\,km\,s$^{-1}$}}
\newcommand{\cmsq}{\mbox{\,cm$^{-2}$}}

\newcommand{\nh}{\mbox{${N}_{\rm H}$}} 
\newcommand{\mrk}{Mrk~231}
\newcommand{\clover}{H~1413+117}
\newcommand{\apm}{APM~08279+5255}
\newcommand{\rxj}{RX~J0911.4+0551}
\newcommand{\pgone}{PG~1115+080}
\newcommand{\pgfour}{PG~1411+442}
\newcommand{\pgfive}{PG~1535+547}
\newcommand{\pgbal}{PG~2112+059}
\newcommand{\phl}{PHL~5200}
\newcommand{\asca}{{\emph{ASCA}}}
\newcommand{\chandra}{{\emph{Chandra}}}

\newcommand{\rosat}{{\emph{ROSAT}}}

\newcommand{\aox}{$\alpha_{\rm ox}$}

\begin{document}
 
 
\shortauthors{Gallagher et al.}
\shorttitle{X-ray Spectroscopy of QSOs with UV BALs}


\title{X-ray Spectroscopy of QSOs with Broad Ultraviolet Absorption Lines}

\author{S.~C.~Gallagher, W.~N.~Brandt, G.~Chartas, and G.~P.~Garmire}
       
\affil{Department of Astronomy and Astrophysics \\ 
       The Pennsylvania State University \\
       University Park, PA 16802 \\ 
	USA \\
       {\em gallsc, niel, chartas, garmire@astro.psu.edu}}

\begin{abstract}
For the population of QSOs with broad ultraviolet absorption lines, 
we are just beginning to accumulate X-ray observations with enough
counts for spectral analysis at CCD resolution. 
From a sample of eight QSOs [including four Broad Absorption Line (BAL)~QSOs
and three mini-BAL~QSOs] with \asca\ or \chandra\ spectra with more than
200~counts, general patterns are emerging.
Their power-law X-ray continua are
typical of normal QSOs with $\Gamma\approx2.0$, and 
the signatures of a significant column
density [\nh\,$\approx(0.1$--$4)\times10^{23}$\cmsq] of intrinsic, absorbing
gas are clear.  Correcting the X-ray spectra for intrinsic absorption recovers 
a normal ultraviolet-to-X-ray flux ratio, indicating that the spectral 
energy distributions of this population are not inherently anomalous. 
In addition, a large fraction of our sample shows significant evidence for
complexity in the absorption. The subset of BAL QSOs with broad \MgII\
absorption apparently suffers from Compton-thick absorption completely obscuring the direct
continuum in the 2--10~keV X-ray band, complicating any measurement of their
intrinsic X-ray spectral shapes. 
\end{abstract}
\keywords{galaxies: active --- quasars: absorption lines --- X-rays: galaxies}

\section{Introduction}
\label{sec:intro}

Since the first surveys with \rosat, Broad Absorption Line (BAL) 
QSOs have been known to have
faint soft X-ray fluxes compared to their optical fluxes
(Kopko, Turnshek, \& Espey 1994; Green \& Mathur 1996).  
\nocite{KoTuEs1994,GrMa1996}
Given the extreme
absorption evident in the ultraviolet (UV), this soft X-ray faintness was
assumed to result from intrinsic absorption.  Based on this model, the 
intrinsic column densities required to suppress the X-ray flux, assuming a
normal QSO spectral energy distribution, were found to be
$\gtrsim5\times10^{22}$\cmsq\ \citep{GrMa1996}.  Due to the 2--10~keV
response of its detectors, a subsequent \asca\
survey was able to raise this lower limit for some objects by an order
of magnitude, to $\gtrsim5\times10^{23}$\cmsq\ \citep{GaEtal1999}.
In all of these studies, the premise of a typical underlying QSO
spectral energy distribution and X-ray continuum was maintained.  The
strong correlation found by Brandt, Laor, \& Wills (2000; hereafter BLW)\nocite{BrLaWi2000} between
\CIV\ absorption equivalent width (EW) and faintness in soft X-rays further 
supported this assumption.

\asca\ observations of two BAL~QSOs, \phl\
\citep{MaElSi1995} and \mrk\ \citep{Iw1999,Tu1999}, provided suggestive 
evidence that intrisic absorption was in fact to blame for X-ray
faintness.  However, the limited photon statistics and bandpass
precluded a definitive diagnosis.  The observation
of \pgbal\ with \asca\ provided the first solid evidence
for intrinsic X-ray absorption and a normal underlying X-ray continuum 
in a BAL~QSO \citep{GaEtal2001a}.
Subsequently, a few more observations of BAL~QSOs with \asca\ and  \chandra\
have upheld this result \citep{GrEtal2001,MaEtal2001,GaEtal2001b}.
Observations of three mini-BAL~QSOs,{\footnote{Mini-BAL~QSOs contain
UV absorption lines with velocity widths of
$\sim10^{3}$~\kms, but they do not formally meet the
BAL~QSO criteria established by \citet{WeMoFoHe1991}.}} \pgfour\ \citep{BrWaMaYu1999},
\rxj\ \citep{ChEtal2001}, and \pgone\ \citep{GaBrChGa2001},
were also consistent with the absorption scenario.  

In this paper, we have gathered the results from the 0.5--10~keV spectral analyses
of all the available QSOs with broad UV absorption lines, including
 BAL~QSOs, mini-BAL~QSOs, and one Narrow-Line Seyfert~1 galaxy, with
more than 200~counts.  Though additional data exist in the literature, these are the only
sources with enough counts for spectral analysis on an individual
basis.  From our experience, drawing strong conclusions about the
nature of an X-ray spectrum based on fewer counts
can lead to interpretations that are not upheld by higher quality data.

\section{Sample Data}
\label{sec:obs}

Our sample is listed in Table~1, along with some relevant
observational parameters. The X-ray data were taken by the \asca\ and \chandra\ observatories,
as indicated in the ``Notes'' column to Table\,1, and 
Figure~\ref{fig:spectra} shows those spectra not previously
published. The analyses for the sample objects are described in the
references listed in Table\,1, with the exceptions of \pgfour\ \citep{BrWaMaYu1999} and
\phl\ \citep{MaEtal2001}.  The spectral analyses for these two objects have been redone
according to the method detailed in \citet{GaEtal2001a} to ensure
consistency with the other results.  For \phl, corrective measures 
for dealing with the degradation of the \asca\ CCD detectors, as
outlined in Weaver, Gelbord, \& Yaqoob (2000; Appendix
A)\nocite{WeGeYa2001}, were followed.  Our derived photon index for
the underlying X-ray power-law continuum pf \phl,
$\Gamma=2.17^{+0.60}_{-0.47}$, was significantly flatter than 
that of the preferred model of \citet{MaEtal2001}.

The general strategy for all of the X-ray analysis was to first try to
fit the data with a simple
power-law model (including Galactic absorption) and subsequently
add intrinsic absorption of three sorts: neutral, partially covering neutral,
and ionized.  In all cases, including intrinsic absorption provided
statistically acceptable models and improved the $\chi^2$ of the fits
significantly.  For the gravitational lenses,
\apm, \rxj, \pgone, and \clover, the lensing galaxies could provide
some of the observed absorption.  However, this contribution is unlikely to be
significant. At the redshifts of the known lenses, the column density
would have to be $\gtrsim10^{22}$~\cmsq, much larger than the typical \nh\ through the
outer regions of a galaxy, to affect strongly the best-fitting column density. 
In general, distinguishing between a partially 
covering and ionized absorber is not possible statistically; with CCD
resolution and the current statistics, the two models can fit the data
equally well. Typically, the ionized-absorber models produce column densities
within a factor of 0.5--2.0 of the partial-covering models.
However, though an ionized absorber was included for
completeness in fitting each of the spectra, the ionized-absorber
models, such as {\textsf{absori}} in the spectral
fitting package {\textsf{XSPEC}} \citep{Ar1996}, assume a single-zone, 
optically thin plasma.  Such assumptions are not in general
appropriate for the large X-ray column densitities,
\nh$\gtrsim10^{22}$\cmsq, exhibited by QSOs with broad UV absorption
(H.~Netzer, 2001, priv. comm.).  Therefore, when a complex absorber model
provided a statistically significant improvement over a simple neutral absorber, the
partial-covering model was chosen.

The column densities and covering fractions for the best-fitting
absorption models, as well as the best-fitting photon indices, are
listed in Table\,1.

\section{Results}

As obtaining even moderate-quality spectra of these faint X-ray sources requires a
significant amount of observing time with the current generation
of X-ray observatories, a
comprehensive and intricate X-ray picture of the population of QSOs with
broad UV absorption lines
awaits the accumulation of a significantly larger body of data.
However, as the number of these QSOs observed in X-rays
grows, a consistent basic picture is beginning to emerge.  

\subsection{Normal Underlying QSO Continua}

\begin{figure*}[t]
\centerline{\includegraphics[scale=0.60]{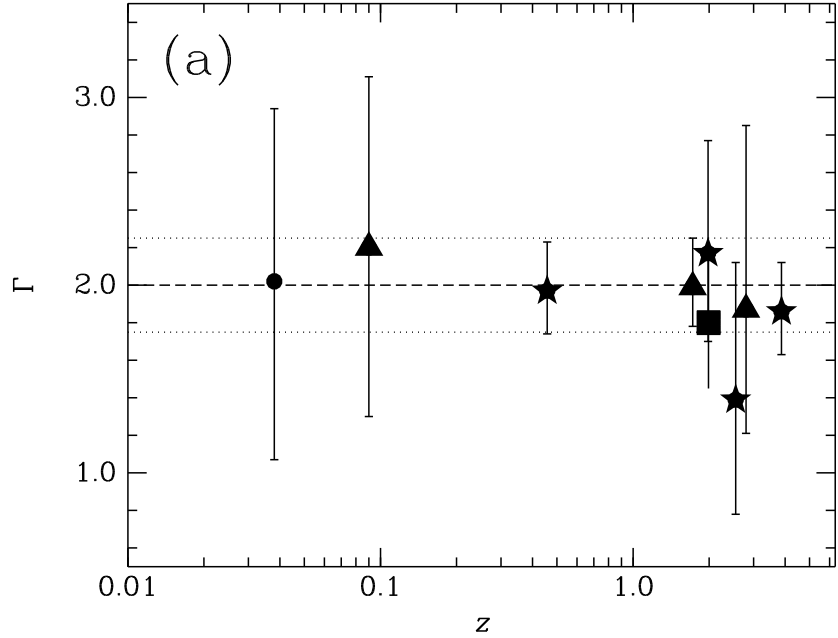}}
\caption{
For all plots, data points are for the QSOs listed in Table~1 unless otherwise indicated.
Filled stars mark BAL~QSOs, filled triangles mark mini-BAL~QSOs,
and the filled circle marks the NLS1, \pgfive.
{\bf (a)} Photon index versus redshift for the QSOs listed in 
Table~1.  The filled square is the result from \citet{GrEtal2001}
for a simultaneous fit to six BAL~QSOs versus their median $z$.
The dashed line represents the mean of the \citet{GeEtal2000}
\asca\ sample of radio-quiet QSOs, and the dotted
lines indicate the dispersion.
{\bf (b)} Histograms of \aox\ values.  The dotted histogram shows the
distribution of the BLW sample of low-redshift Palomar-Green QSOs
without known X-ray absorption. The shaded histogram shows the
distribution of \aox\ for our sample as listed in Table~1.  The
filled symbols show $\alpha_{\rm ox(corr)}$, the \aox\ values for our sample
corrected for intrinsic X-ray absorption.
{\bf (c)} Intrinsic X-ray column density versus rest-frame \CIV\
absorption EW.
\label{fig:plots}
}
\end{figure*}

Most fundamentally, the days of assuming, without evidence, that
QSOs with broad UV absorption lines have the underlying X-ray
continua of normal QSOs are over.  
The spectroscopic measurement of the photon index, $\Gamma$, of the power-law
continuum for each of these objects versus redshift is shown in
Figure\,\ref{fig:plots}a. 
Excluding the apparent outlier \clover, the mean photon index for our sample is
$\Gamma=2.01\pm0.13$, entirely consistent with the range of
$\Gamma$ measured for large samples of 
low and high luminosity radio-quiet QSOs \citep[e.g.,][]{GeEtal2000,ReTu2000}.
Furthermore, there is no evidence for strong systematic changes in spectral
slope as a function of redshift.  
In fact, the well-measured continuum slope of \apm\
($\Gamma=1.86^{+0.26}_{-0.23}$) extends constraints on
radio-quiet QSO X-ray continuum shape evolution to significantly
higher redshift, $z=3.87$, than previously possible.

For evaluating the spectral energy distribution of a QSO, the quantity
\aox\ measures the relative
flux densities at 2~keV and 3000~\AA.  Radio-quiet QSOs without
intrinsic absorption have \aox\,$=-1.51\pm0.14$ (BLW), while absorbed QSOs tend to have
\aox\,$\lesssim-1.7$.  The \aox\ values for this sample are found to be
generally indicative of absorption (see Table~1 and Figure~{\ref{fig:plots}b}),
though the measured X-ray column-density is not obviously correlated
with \aox. Though our values for the 3000~\AA\ flux density
were not corrected for possible intrinsic reddening, this is not
expected to be a significant problem for these blue QSOs.
Furthermore, though variability 
between the optical and X-ray observations could result in an
inaccurate \aox, this quantity is only sensitive to large changes in
flux density; e.g., a factor of two increase in the 3000~\AA\
flux density only corresponds to $\Delta$\aox$=0.1$.
Correcting the X-ray spectra for the fitted intrinsic absorption
brings these objects within the range of typical \aox, indicating that
the underlying spectral energy
distributions are not anomalous (see Figure~{\ref{fig:plots}b}). Both 
of the assumptions of the early BAL~QSO X-ray work, normal X-ray
continuum and typical \aox, are thus supported.

\subsection{Complex Absorption}

A simple neutral absorber blocks almost all X-ray flux up to an energy
set by the column density; for \nh$=10^{23}$\cmsq, X-rays up to
\mbox{$\sim3$\,keV} in the rest frame are absorbed.  A signature of
more complex absorption, for example partial covering and/or ionized gas 
along the line of sight, is a much less abrupt decline of the flux at the lowest
energies (see \mbox{Figure\,\ref{fig:spectra}b} for an example).
For five of the eight sources, a partial-covering absorber provided a statistically
significant improvement over a neutral absorber, and the column
densities listed are for that model.

For the remaining three QSOs, \apm, \clover, and \pgbal, the current data do not
permit the nature of the absorber to be examined in detail.  
For \clover\ and \pgbal, the
signal-to-noise ratio of the X-ray data below \mbox{1\,keV} is
insufficient for studying the structure of the absorption, and for
\apm, the large redshift ($z=3.87$) pushes the signatures of a complex
absorber below the \chandra\ ACIS bandpass (\mbox{0.5--8.0\,keV}).

\subsubsection{\clover}
\label{sec:clover}

The famous gravitationally lensed ``Cloverleaf'' QSO, \clover, does not fit
within the basic picture outlined above for the other absorbed QSOs in this sample.  Though
significant intrinsic absorption is measured,
the best-fitting photon index is unusually flat for a radio-quiet QSO,
$\Gamma\approx1.4$. Furthermore, even after correcting for intrinsic
absorption, \aox\ remains low at $-1.74$.
Finally, including a narrow, 6.4~keV \feka\ emission line in the model improves 
the fit at the $>99\%$ confidence level according to the $F$-test
($\Delta\chi^2=9.1$ for one additional fitting parameter;
\mbox{Figure\,\ref{fig:spectra}c}).  In the rest frame, this strong line has
an equivalent width of ${\rm EW}=650^{+354}_{-517}$~eV.
The flat rest-frame 1.8--18~keV slope and strong \feka\ emission line together suggest
that the spectrum is reflection-dominated, and thus the
direct X-rays are completely obscured.  In terms of absorption-line
profiles, \clover\ is similar to \phl, with smooth profiles starting
at near-zero velocities \citep{TuGrFoWe1988}, and both QSOs also exhibit
high continuum polarization \citep{GoMi1995}.  However, the X-ray
properties of these two QSOs are quite different as \phl\ shows no
evidence for reflection. The difference between 
\clover\ and the other BAL~QSOs may lie in its UV spectrum; 
\citet{HaMoTeMc1984} found broad \AlIII\ absorption, which has never been
found without \MgII\ absorption, the signature of a low-ionization BAL~QSO
(LoBAL~QSO; N. Arav and M. Brotherton 2001, priv. comm.).

Those BAL~QSOs with broad absorption in \MgII\ have notably lower \aox\ values 
than BAL~QSOs with only
high-ionization absorption lines \citep{GaEtal1999,GrEtal2001}.
This could plausibly result from either an intrinsically
X-ray weak continuum, or from heavy, perhaps Compton-thick,
obscuration \citep[e.g.,][]{GrEtal2001}.  In the latter case, only
indirect X-rays from the active nucleus, either scattered around the
obscuration by a small-scale electron mirror or reflected off of more
distant, neutral material, can reach the observer, and the intrinsic
X-ray power of the nucleus cannot be directly measured.
Given the extreme X-ray faintness of this class of objects, very few
spectra are likely to be obtained.  The two examples available at
present are Mrk~231 \citep{MaRe2000,GaEtal2001b}
and \clover.  For these two objects, two pieces of evidence
indicate that the direct X-ray continuum is not being
observed: the spectral shape is unusually flat ($\Gamma\lesssim1.5$),
and the value of \aox\ is still low after correcting for all the
direct signs of absorption.

\section{Discussion}

In every case where the analysis is sensitive to structure in the
X-ray absorption, we find evidence for complexity.  While we model this
complexity as intrinsic, neutral absorption partially covering the
X-ray continuum source, the actual nature of this absorbing gas could be much more complicated.
X-ray scattering in the nucleus effectively causes partial-covering
absorption, and if the X-ray absorbing gas is physically related to that absorbing the
UV continuum, then it should also be ionized. Determining the ionization state of the
X-ray absorbing gas and the continuum scattering fraction is essential 
for accurately measuring the column density.  Furthermore, the current
modeling assumes no velocity dispersion in the absorbing gas. In such
gas, the opacity in the soft X-rays arises primarily from
bound-free metal edges, and these are used to determine the absorption
column density.  With a significant velocity dispersion, the opacity of bound-bound
absorption lines can increase significantly.  Therefore, the assumption of zero
velocity dispersion, clearly na\"{\i}ve for BAL~QSOs, may result
in an overestimate of the amount of X-ray absorbing gas. 

The significant correlation found by BLW of \CIV\
absorption EW versus weakness in soft X-rays holds across several
orders of magnitude of absorption EW.  The inference that the weakness in soft X-rays
is the result of intrinsic absorption has been substantiated
spectroscopically for all but one of the objects in our sample
(\clover, see $\S$\,\ref{sec:clover}).  However, directly identifying the EW 
of the UV absorption with the column density of the X-ray
absorbing gas is problematic (see \mbox{Figure\,\ref{fig:plots}c}).  BLW
speculated that at the absorbed end of their correlation, the EW of
\CIV\ is dominated primarily by velocity dispersion rather than column 
density.  In this case, BAL~QSOs, by definition \citep{WeMoFoHe1991},
would be expected to have the most extreme \CIV\ EW properties, as
they do.  As seen in \mbox{Figure\,\ref{fig:plots}c} though,
BAL~QSOs may not necessarily always have the largest X-ray absorption
column densities.

In terms of gross X-ray properties, it has become apparent that
drawing a distinction between BAL QSOs and other QSOs with broad UV absorption lines is 
largely a matter of semantics.  The photon indices and the column
densities measured with the current X-ray observatories overlap. Future, higher spectral
resolution X-ray observations may reveal features warranting
separate classifications, but at present the differences do not justify
treating the BAL and mini-BAL QSO populations, for example, as distinct.  

\section{Future Work}

Though X-ray spectroscopy of QSOs with broad UV absorption lines is finally
addressing some of the questions first raised by \rosat\
observations, many issues remain unresolved.  Primarily, the velocity
structure and ionization state of the X-ray absorbing gas are unknown.
Without this crucial information, the mass ejection rate of gas from
the nuclei of these QSOs cannot be determined.  This parameter is
fundamental for understanding the balance of
accretion and outflow in the central engine.
To address these important questions, X-ray spectroscopy at gratings
resolution of the X-ray brightest QSOs with broad UV absorption is the next step.
Investigating the physical basis for the extreme X-ray faintness of
LoBAL~QSOs may need to await the next generation of X-ray observatories.

\acknowledgements
We thank F. Hamann for providing equivalent-width data, C. Vignali for
helpful discussions, and M. Brotherton for constructive comments that
improved this paper. NASA grant NAS 8-38252, P.I. GPG, supports the ACIS instrument team.
SCG gratefully acknowledges support from NASA GSRP grant NGT5-50277 and
from the Pennsylvania Space Grant Consortium. WNB thanks NASA 
LTSA grant NAG5-8107.


\clearpage

\begin{deluxetable}{llccccccl}
\rotate
\tablewidth{0pt}
\tablecaption{Basic Properties of QSOs with Broad Ultraviolet Absorption Lines
\label{tab:1}}
\tablehead{
\colhead{}	&
\colhead{}	&
\colhead{}	&
\colhead{Intrinsic \nh\tablenotemark{b}}	&
\colhead{Covering}	&
\colhead{}	&
\colhead{}	&
\colhead{\CIV\ EW\tablenotemark{d}}	&
\colhead{}	\\
\colhead{Name\tablenotemark{a}}	&
\colhead{$B$}	&	
\colhead{$z$}	&
\colhead{($10^{22}$\cmsq)}	&
\colhead{Fraction\tablenotemark{b}}	&
\colhead{$\Gamma$\tablenotemark{b}} &
\colhead{\aox/$\alpha_{\rm ox(corr)}$\tablenotemark{c}} &
\colhead{(\AA)}	&
\colhead{Notes\tablenotemark{e}} \\
}
\startdata
\apm$^{1}$	& 19.2	& 3.870	& $6.0^{+3.5}_{-3.1}$
&$\cdots$	& $1.86^{+0.26}_{-0.23}$ & $-2.26/-1.70^{5}$	& 21$^{5}$ & B, GL, C	\\
\rxj$^{2}$ 	& 18.0  & 2.800 & $19^{+28}_{-18}$
&$0.71^{+0.20}_{-0.39}$	& $1.87^{+0.98}_{-0.66}$ & $-1.72/-1.52^{6}$
& 4.4$^{6}$	& mB, GL, C\\ 
\pgone$^{1}$	& 15.8  & 1.722 & $3.8^{+2.5}_{-2.2}$
&$0.64^{+0.12}_{-0.17}$	& $1.99^{+0.26}_{-0.21}$ & $-1.74/-1.60^{7}$	& $\cdots$	& mB, GL, C\\
\clover$^{1}$	& 17.0  & 2.548	& $20^{+18}_{-12}$     
&$\cdots$	& $1.39^{+0.73}_{-0.61}$ & $-2.17/-1.74^{8}$	& 46$^{9}$	& B, GL, C\\	
\pgfour$^{3}$	& 15.0  & 0.090 & $19^{+7.5}_{-6.6}$
&$0.97^{+0.02}_{-0.12}$	& $2.20^{+0.91}_{-0.90}$ & $-2.01/-1.44^{7}$	& 8.5$^{10}$	& mB, A	\\
\pgfive$^{4}$	& 15.3  & 0.038 &  $12^{+8}_{-5}$
&$0.91^{+0.07}_{-0.26}$	& $2.02^{+0.92}_{-0.95}$ & $-2.03/-1.66^{7}$	& 4.6$^{10}$  	& NLS1, A\\
\pgbal$^{4}$ 	& 15.5	& 0.457	& $1.1^{+0.5}_{-0.4}$	
&$\cdots$	& $1.97^{+0.26}_{-0.23}$ & $-1.62/-1.55^{7}$	&
19$^{10}$	& B, A	\\	
\phl$^{3}$	& 18.0	& 1.981 &  $40^{+26}_{-22}$
&$0.79^{+0.12}_{-0.23}$	& $2.17^{+0.60}_{-0.47}$ & $-1.66/-1.44^{8}$	& 55$^{9}$	& B, A	\\	
\enddata
\tablenotetext{a}{Numerical superscripts in this column indicate the
reference for the detailed description of the X-ray
analysis.  $^{1}$Chartas et al., in prep.; $^{2}$\citet{ChEtal2001};
$^{3}$see \S\ref{sec:obs}; $^{4}$\citet{GaBrChGa2001}. }
\tablenotetext{b}{Errors are given for $90\%$ confidence taking
all parameters except normalization to be of interest.}
\tablenotetext{c}{The values for $\alpha_{\rm ox(corr)}$ were calculated 
from the X-ray spectrum corrected for intrinsic absorption. Numerical
superscripts indicate the references for the 
3000~\AA\ flux densities used to calculate \aox\ and $\alpha_{\rm ox(corr)}$.
 $^{5}$\citet{ElEtal1999}; $^{6}$\citet{BaEtal1997}; $^{7}$Neugebauer
et al. (1987); $^{8}$Weymann et al. (1991).}
\tablenotetext{d}{Sources of \CIV\ absorption EW values are indicated
with numerical superscripts. $^{9}$\citet{HaKoMo1993}; $^{10}$\citet{BrLaWi2000}.}
\tablenotetext{e}{Key: B: BAL~QSO; mB: mini-BAL~QSO; NLS1: Narrow-Line
Seyfert~1; GL: gravitational lens.  The sources of the data are \asca\ (A) and 
\chandra\ (C).}
\end{deluxetable}
\clearpage
\begin{figure*}[t]
\centerline{\includegraphics[scale=0.4,angle=-90]{f1a.eps}}
\centerline{\includegraphics[scale=0.4,angle=-90]{f1b.eps}}
\centerline{\includegraphics[scale=0.4,angle=-90]{f1c.eps}}
\caption{
Observed-frame \chandra\ ACIS-S3 spectra.  Both  \textbf{(a)} \apm\ and
 \textbf{(b)} \pgone\ have been fit above rest-frame \mbox{5\,keV}
with a power-law model that has then been extrapolated back to lower
energies.  The residuals at low energies are signatures of significant 
intrinsic absorption.
\textbf{(c)}
\clover. The continuum has been fit with a power-law model with
intrinsic absorption. Significant positive residuals near
rest-frame 6.4\,keV are indicative of a strong neutral
\feka\ emission line with rest-frame \mbox{EW$\approx650$\,eV}.
\label{fig:spectra}
}
\end{figure*}

\clearpage

\begin{figure*}[t]
\figurenum{2b and 2c}
\centerline{\includegraphics[scale=0.60]{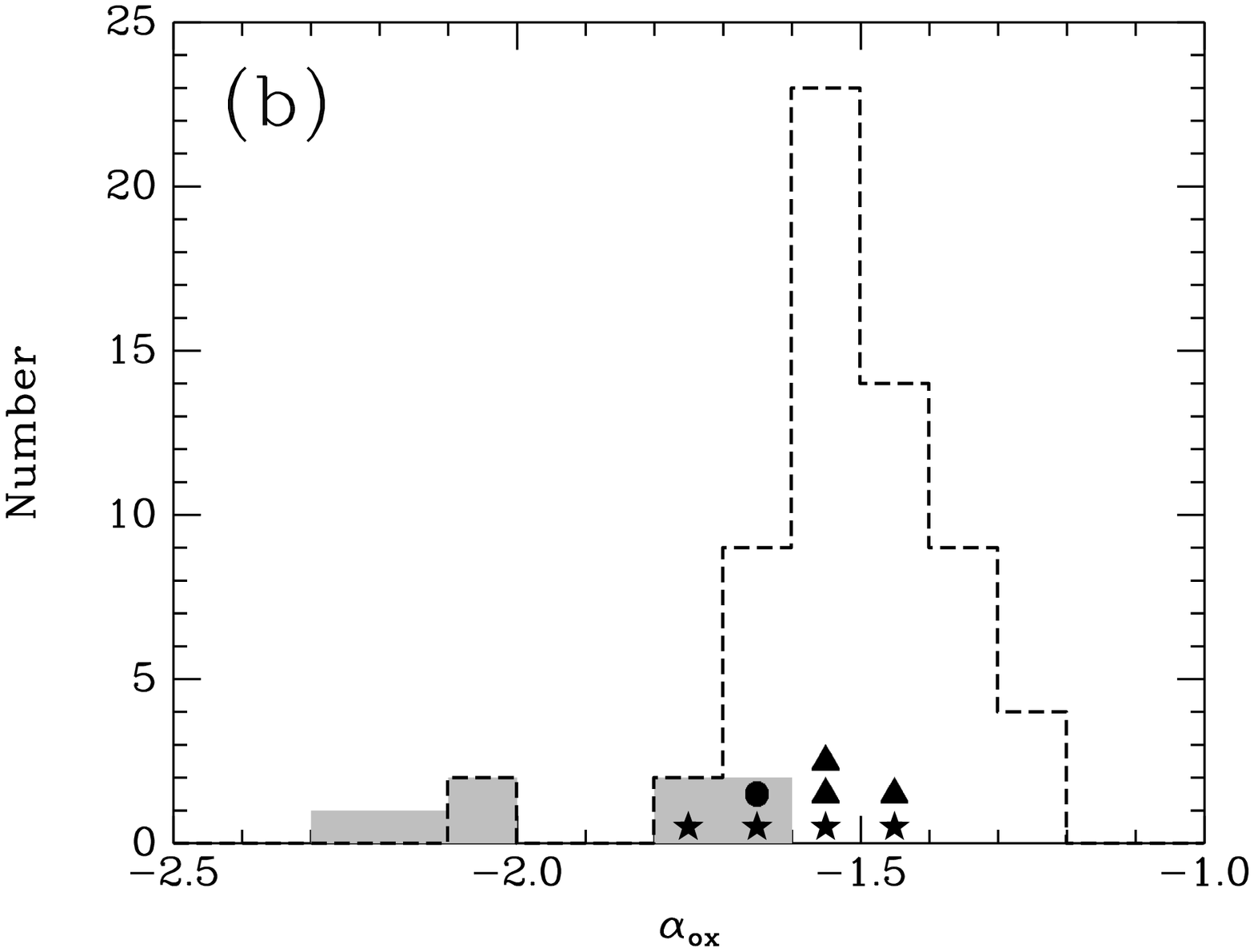}}
\centerline{\includegraphics[scale=0.60]{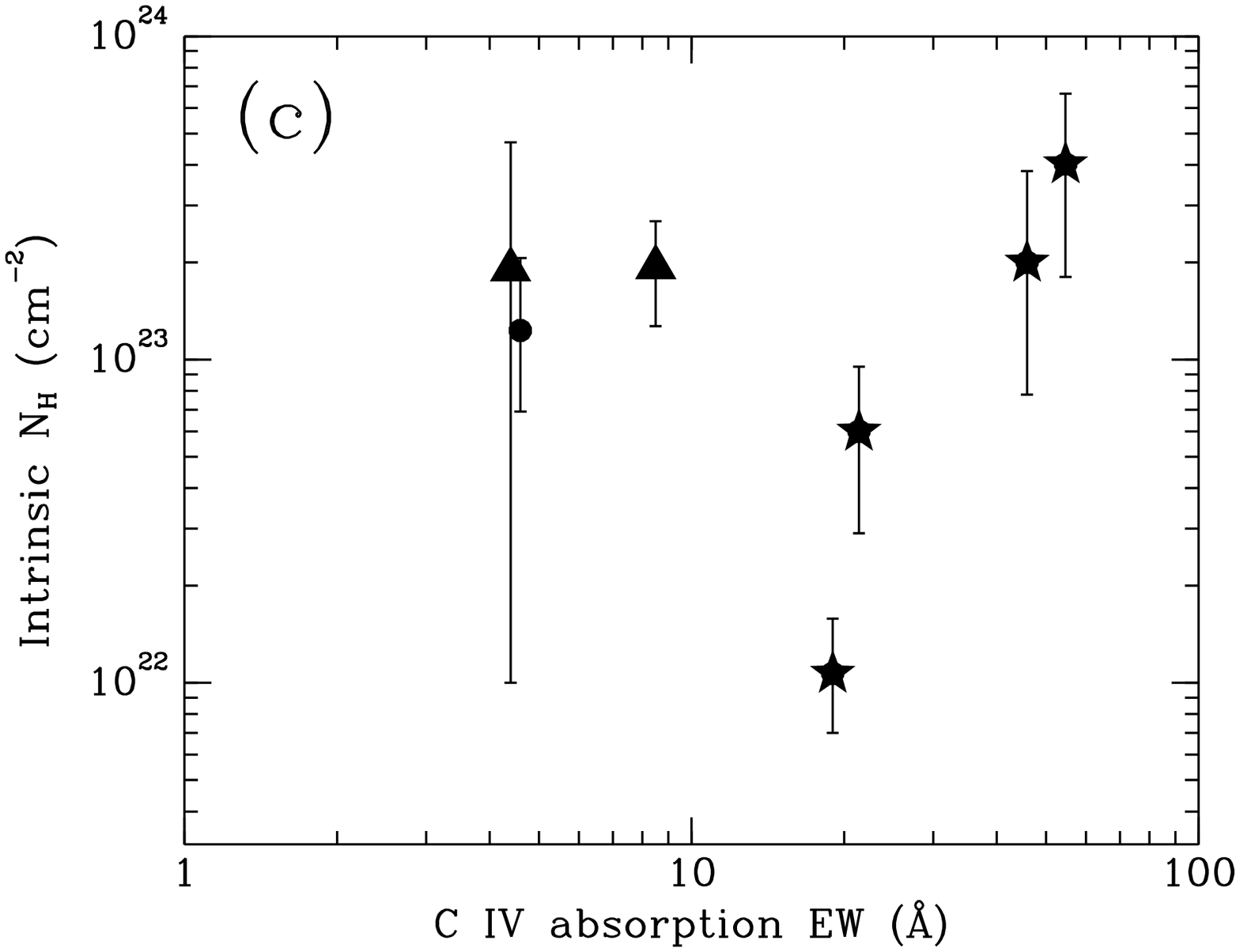}}
\end{figure*}

\end{document}